# Bulk magnetization and strong intrinsic pinning in Ni-doped BaFe$_2$As$_2$ single crystals

K S Pervakov[1,2,*], V A Vlasenko[1], E P Khlybov[2,3], A Zaleski[4], V M Pudalov[1] and Yu F Eltsev[1]

[1]P. N. Lebedev Physical Institute, Russian Academy of Sciences, Moscow 119991, Russian Federation;

[2]International Laboratory of High Magnetic Fields and Low Temperatures, Wroclaw 53-421, Poland;

[3]Institute for High Pressure Physics, Russian Academy of Sciences, Troitsk 142190, Moscow region, Russian Federation;

[4]Institute of Low Temperature and Structure Research, Polish Academy of Sciences, P.O. Box 1410, 50-422 Wroclaw, Poland

*Corresponding author (pervakov@sci.lebedev.ru)

We report on measurements of isothermal irreversible magnetization loops *(M(H))* of Ni-doped BaFe$_{2-x}$Ni$_x$As$_2$ single crystals with *x*=0.1 and *x*=0.14 in magnetic fields up to 14T. For both samples in field *H*//*c*-axis and *H*//*ab*-plane, critical current density calculated from *M(H)* curves exceeds $10^6$ A/cm$^2$ at low temperatures suggesting strong intrinsic pinning in our crystals. In a broad temperature range 2-17K in both field orientations nearly optimally doped single crystal displays second magnetization peak (fish-tail effect). For *H*//*c*-axis, curves of the normalized pinning force, $f_p$, as a function of reduced field, $h$, measured at different temperatures obey scaling relation $f_p \propto h^p(1-h)^q$ with peak position $h_{max} \approx 0.33$ for BaFe$_{1.86}$Ni$_{0.14}$As$_2$ crystal and $h_{max} \approx 0.4$ for BaFe$_{1.9}$Ni$_{0.1}$As$_2$ sample indicating single normal point pinning mechanism. In striking contrast, in H//*ab*-plane orientation, the scaling of $f_p(h)$ curves is absent.

**Introduction**

Already first studies of recently discovered iron based superconductors (FBS) revealed high potential of these compounds for strong magnetic field applications. In particular, band structure calculations as well as angular-resolved photoemission spectroscopy (ARPES) measurements showed low Fermi velocities, $v_F$, of the order of $10^6$-$10^7$ cm/sec resulting in extremely short coherence length $\xi \sim \hbar v_F / 2\pi k_B T_c \sim$1-3nm and, consequently, very high values of the upper critical field $H_{c2} = \phi_0 / 2\pi \xi^2$ exceeding 100T[1]. Similar estimation of $H_{c2}(0)$ has been obtained from direct experimental measurements of $H_{c2}(T)$ value in accessible magnetic fields and in extrapolation to zero temperature[2-4]. The most intensively studied FBS samples of the 122 family in addition to moderately high superconducting critical temperature, $T_c$, up to 30-35K, demonstrate high critical current density, $J_c$, in the range of $10^5$-$10^6$ A/cm$^2$ at T=4.2K[5-14] and relatively low anisotropy $\gamma = H_{c2}^{(ab)} / H_{c2}^{(c)}$=1-2[3,10]. Furthermore, in contrast to the cuprate high-$T_c$ superconductors Weiss *et al*[15] have found evidence for high intergrain $J_c$ in fine grain $Ba_{0.6}K_{0.4}Fe_2As_2$ samples.

Further understanding of the origin of strong intrinsic pinning in FBS compounds and details of the vortex dynamics in magnetic field directed both perpendicular and parallel to the *ab*-plane is an important stepping stone to development of FBS samples with high superconducting current carrying ability. In recent studies of several research groups a few general features of the vortex dynamics in FBS of the 122 family have been reported: (i) the second magnetization peak (fish-tail effect) was observed in nearly optimally doped single crystals of different composition with electron as well as hole doping.[5,6,8-11,13,14,16-18] To account for the second magnetization peak in FBS a few different mechanisms have been suggested (see Ref. 16 and references therein) but still unified point of view on the nature of the fish-tail anomaly is absent; (ii) a temporal magnetic relaxation with typical magnetization versus time dependence, $M \propto log(t)$ was reported,[6-8,13,16,17] whereas in Ref.18 anomalous relaxation behavior showing two distinct time windows of the logarithmic $M \propto log(t)$ dependence with different slopes has been observed; (iii) a crossover was found from collective elastic to plastic creep at magnetic field dependent temperature identified from $J_c(T)$ measurements and/or the temperature dependence of the flux creep rate,[5-8,13,16] and finally (iv) in magnetization measurements with rather high field sweep rate (~100 Oe/s) there were observed in 122 single crystals[11-13] magnetic flux jumps, that are usually associated with strong pinning strength and high critical current density.

Optimally doped $BaFe_{2-x}Ni_xAs_2$ single crystals have lower $T_c$ and $H_{c2}(0)$ compared to $Ba_{1-x}K_xFe_2As_2$ and $BaFe_{2-x}Co_xAs_2$ systems of 122 family with maximal superconducting critical temperature approaching 40K and around 25K, correspondingly.[14] Probably for this reason Ni-doped 122 crystals were studied not so intensively like K- and Co-doped 122 compounds. On the other hand there is a nice opportunity to study superconducting current caring capacity and vortex pinning in Ni-doped $BaFe_2As_2$ single crystals over wide temperature range covering almost all magnetic phase diagram of this compound. In this paper we report on measurement of isothermal bulk magnetization loops of $BaFe_{2-x}Ni_xAs_2$ single crystals with two levels of doping: nearly optimally doped ($x$=0.1, $T_c \approx$ 19.5K) and slightly overdoped ($x$=0.14, $T_c \approx$ 13K) in magnetic fields up to 14T applied perpendicular and parallel to the *ab*-plane. For optimally doped sample we observed second magnetization peak in the broad temperature range (2÷17K) for both field orientations and $J_c$ value approaching $10^7$ A/cm$^2$ at low temperatures. For both samples the curves of normalized pinning force $f_p = F_p/F_p^{max}$ *versus* reduced field $h = H/H_{irr}$, (where $H$ - irreversibility field), measured at different temperatures collapse into a single curve for H//*c*-axis field orientation. This is in striking contrast to H//*ab*-plane geometry where scaling of the $f_p(h)$ curves is absent. Scaling of the $f_p(h)$ curves in the H//*c*-axis configuration indicates a single dominating vortex pinning mechanism.

**Experiment and discussion**

$BaFe_{2-x}Ni_xAs_2$ single crystals with $x$=0.1 and $x$=0.14 were grown using the self-flux method. Starting components of high purity Ba, FeAs and NiAs of total amount of ~5g were mixed in 1:5(1-x):5x molar ratio, placed in alumina crucible of about 3cm$^3$ volume, sealed in quartz tube under 0.2 atm of argon pressure and loaded into a tube furnace. Next, the ampoule was heated up to 1200°C, held at this temperature for 24 hours for homogenization melting, and then cooled down to 1070°C with 2°C/h rate. At this temperature the ampoule together with a furnace was turned over from vertical to horizontal position to decant liquid flux from the crystals. Further, the ampoule with crystals was cooled down to room temperature inside the furnace. Finally, crystals of size up to 4x2 mm$^2$ in the *ab*-plane were collected from the crucible. Characterization of crystals using x-ray diffraction showed absence of any foreign phases for crystals of both compositions with Ni doping level *x*=0.1 and *x*=0.14. For measurements we cleaved grown crystals to 0.1÷0.2 mm thickness and cut them into rectangular shape. The mass of BaFe$_{1.9}$Ni$_{0.1}$As$_2$ crystal was about 56 mg, for another sample BaFe$_{1.86}$Ni$_{0.14}$As$_2$ it was approximately 13 mg. Temperature dependence of ac-susceptibility was obtained using the Quantum Design PPMS system. To measure irreversible magnetization loops we used home-built low-frequency (3.6 Hz) vibrating sample magnetometer with a step motor.[19] Typical field sweep rate was inside the range 20-90 Oe/s.

High quality of our BaFe$_{2-x}$Ni$_x$As$_2$ single crystals is well illustrated by Fig.1 where we show temperature dependence of the real, $\chi'$, and imaginary, $\chi''$, part of ac-susceptibility for samples with *x*=0.1 (top panel) and *x*=0.14 (bottom panel) measured in zero field and in magnetic fields of 1, 5 and 9 T applied along the *c*-axis. One can see zero field superconducting transition width less than 2 K and its shift to lower temperatures without broadening in increasing field. Critical temperature of the superconducting transition, $T_c$, defined by extrapolation of a linear part of the transition to zero signal is 19.5K and 13K for crystals with *x*=0.1 and *x*=0.14, correspondingly. Rough estimation of a slope of the upper critical field, $H_{c2}$, as a function of temperature gives $dH_{c2}/dT \approx 4.2$T/K and $\approx 3.6$T/K for BaFe$_{1.9}$Ni$_{0.1}$As$_2$ and BaFe$_{1.86}$Ni$_{0.14}$As$_2$ crystals respectively.

In Fig.2 we present data of bulk magnetization *versus* field applied along the *c*-axis (left panels) and parallel to the *ab*-plane (right panels) for crystals with Ni doping level *x*=0.1 (upper panels) and *x*=0.14 (lower panels) obtained at different temperatures. High symmetry of our *M(H)* irreversible magnetization loops with respect to the external field is clearly seen from Fig.2. In agreement with previously published results, the isothermal *M(H)* hysteresis loops for both our samples show a sharp zero field central peak. Also, for BaFe$_{1.9}$Ni$_{0.1}$As$_2$ single crystal we observed a broad second peak for both magnetic field orientations whereas for BaFe$_{1.86}$Ni$_{0.14}$As$_2$ sample the fish-tail anomaly is absent. Similar to other studies, with increasing temperature the second peak position moves to lower fields. As we already mentioned, the origin of the second peak may be attributed to several mechanisms.[16]

Our results of the *M(H)* hysteresis loops measurements open a possibility to determine with reasonable accuracy irreversibility fields, $H_{irr}^{(c)}$ and $H_{irr}^{(ab)}$ for *H//c*-axis and *H//ab*-plane magnetic field orientations, correspondingly, at different temperatures when hysteresis loop width, $\Delta M = M_{down} - M_{up}$, where $M_{up}$ and $M_{down}$ - magnetization measured with increasing and decreasing field, falls below instrument resolution limit within the range of accessible in our experiment magnetic fields up to 14T. For lower temperatures, $\Delta M$ starts to decrease with increasing field though is still finite at field of 14T (see the data in Fig.2: for sample BaFe$_{1.9}$Ni$_{0.1}$As$_2$ - the curve at T=7K for *H//c*-axis and the curves at T=13K and T=10K for *H//ab*-plane; for sample BaFe$_{1.86}$Ni$_{0.14}$As$_2$ - the curves at T=1.4K for both field orientations). In this case, in order to estimate $H_{irr}^{(c)}$ and $H_{irr}^{(ab)}$ (somewhat arbitrarily) we used an empirical extrapolation to zero of a part of *M(H)* curves below inflection point in the form $M = M_0 + A\exp(C \times H)$, where $M_0, A$ and $C$ are adjustable parameters; this functional form was verified to fit our data at higher temperatures.

Determined in this way values of $H_{irr}^{(c)}$ and $H_{irr}^{(ab)}$ for both our samples in two fields orientations corresponding to the critical current density of the order of 10$^4$A/cm$^2$ are shown in Fig.3 as a function of temperature. In Fig.3 we present also data for $H_{c2}^{(c)}(T)$ obtained in the *H//c*-axis geometry for single crystals using ac-susceptibility measurements. As expected for both samples irreversibility line $H_{irr}^{(c)}(T)$, separating vortex solid from the vortex liquid is located slightly below $H_{c2}^{(c)}(T)$ phase boundary. For both our samples we obtain slightly temperature dependent ratio $H_{irr}^{(ab)}/H_{irr}^{(c)}$ of about 2. Although there is no direct relation between the irreversibility field and the coherence length $\xi$, the ratio $H_{irr}^{(ab)}/H_{irr}^{(c)}$ observed in our study is consistent with previously reported superconducting anisotropy parameter $\gamma = H_{c2}^{(ab)}/H_{c2}^{(c)} \approx$ 1-2 for FBS single crystals of the 122 family.[3,10]

In order to calculate critical current density from irreversible magnetization loops shown in Fig.2 we used the well-known expression $J_c = 20\Delta M/a(1 - a/3b)$ obtained within Bean critical state model[20] where $a$ and $b$ ($b > a$) are crystal sizes in the plane perpendicular to applied magnetic field. In Fig.4 field dependences of critical current density found from these calculations is shown for magnetic field applied along the $c$-axis (left panels) and parallel to the $ab$-plane (right panels) for crystals with Ni doping level $x$=0.1 (upper panels) and $x$=0.14 (lower panels) at different temperatures. As expected from Fig.2, at all temperatures used in our experiment for sample BaFe$_{1.9}$Ni$_{0.1}$As$_2$, the $J_c(H)$ dependence shows non-monotonic behavior with a broad peak moving to lower fields with increasing temperature. In contrast, for BaFe$_{1.86}$Ni$_{0.14}$As$_2$ crystal, $J_c$ as a function of increasing field monotonically goes to zero. For both samples and both field orientations, the $J_c$ value obtained at low temperatures exceeds 10$^6$A/cm$^2$ being at the upper limit of the critical current density of the 122 family single crystals reported before.[5-14]

A powerful tool to analyze the origin of pinning centers governing critical current density dependence of type-II superconductors on applied magnetic field is to plot normalized pinning force $f_p = F_p/F_p^{max} = J_c(H) \times H/(J_c(H) \times H)_{max}$ as a function of the reduced field $h = H/H_{c2}$. However, in case of FBS samples, similar to high-$T_c$ cuprates, the difference between the upper critical field and irreversibility field is sizable. For this reason it looks reasonable to normalize applied field using instead of $H_{c2}$ irreversibility field $H_{irr}$, where $F_p$ and $J_c$ fall to zero. According to Dew-Hughes[21] in case of single vortex pinning, $f_p(h)$ curves obtained at different temperatures obey a scaling relation $f_p \propto h^p(1 - h)^q$ and should collapse in a single curve with peak position giving information about the origin of pinning centers. In particular, peak position at $h_{max}$=0.2 suggests grain-boundary pinning, $h_{max}$=0.33 corresponds to pinning by the normal point defects of size slightly exceeding coherence length $\xi$, peak at $h_{max}$=0.7 is due to pinning caused by the order parameter spatial variations[21]. In a model developed by Kramer[22] dense strong pinning defects produce a high peak at low $h$, whereas weaker and fewer pinning centers will result in a low peak in $F_p(h)$ at higher $h$.

In Fig.5 we present $f_p(h)$ dependence for BaFe$_{1.9}$Ni$_{0.1}$As$_2$ and BaFe$_{1.86}$Ni$_{0.14}$As$_2$ single crystals obtained at various temperatures in two field orientations using the data for $H_{irr}(T)$ from Fig.3 and $J_c(H)$ data at corresponding temperatures from Fig.4. For both our samples in $H$//$c$-axis field geometry all experimental curves within experimental error collapse into a single curve suggesting one single pinning mechanism. Scaling of the $f_p(h)$ dependence for BaFe$_{1.9}$Ni$_{0.1}$As$_2$ crystal found in the broad temperature range 2 - 17 K where we also observed second peak of magnetization allows us to exclude a relationship between fish-tail effect and crossover from elastic collective to single vortex pinning[6,8,16] or phase transition of the vortex state[17] near the second peak.

Furthermore, for BaFe$_{1.86}$Ni$_{0.14}$As$_2$ crystal the scaling curve is well described by $f_p(h) \sim h(1-h)^2$ expression with peak position $h_{max} \approx 0.33$ in line with Dew-Hughes model[21] signaling normal point pinning (NPP) in this sample. For BaFe$_{1.9}$Ni$_{0.1}$As$_2$ sample the best fit was found for function $f_p(h) \sim h^{1.53}(1-h)^{2.25}$ with $h_{max} \approx 0.4$. Following Kramer[22] we can suggest slightly weaker and fewer normal point defects in this crystal compared to BaFe$_{1.86}$Ni$_{0.14}$As$_2$ sample. Our observation of $h_{max}$=0.33 and 0.4 well agrees with experiments on hole- and electron-doped crystals of the 122 family where peak positions $h_{max}$ of 0.32, 0.37 and 0.43 were reported[14], whereas lower value of $h_{max}$=0.28 in Ba$_{0.65}$Na$_{0.35}$Fe$_2$As$_2$ single crystal[13] as well as slightly higher $h_{max} \approx 0.45$ in Ba(Fe$_{1-x}$Co$_x$)$_2$As$_2$ crystals also were noted.[10] The origin of strong intrinsic pinning in single crystal of the 122 family was discussed in various studies.[10,13,14,23] Inhomogeneous distribution of As or dopants (Ni, Co, K, Na) is one of possible origins of dense normal point defects[10,13,14]; other possibilities like directly observed structural domains due to the orthorhombic distortion also cannot be excluded.[23]

To compare our results for BaFe$_{2-x}$Ni$_x$As$_2$ single crystals with other data in more detail we summarize in Table 1 available in the literature values of $T_c$, $dH_{c2}/dT$, $J_c$, and $h_{max}$ obtained by various research groups for 122 single crystals of different compositions. One can see from the Table reasonable scattering (excluding Refs. 6 and 9) of reported $dH_{c2}/dT$ values which may be due to different methods of used measurements (resistivity, ac-susceptibility, magnetization) as well as different criterions to determine superconducting transition temperature in applied field. Listed in the Table values of critical current density were obtained at various temperatures and $J_c$ spread covers about one order of magnitude. Nevertheless, this scattering also does not look very strong taking into account large difference of $T_c$ values of samples with various doping, possible error in estimation of small samples sizes as well as different samples growth procedure (cooling rate during growth of single crystals and reaching room temperature after that, amount and purity of starting components, chemical composition of crucible, etc.) that may contribute to the density of pinning centers in different samples. In our opinion, the most general feature of observations presented in the Table is scaling of normalized pinning force as a function of reduced field for single crystals with various doping demonstrating single vortex pinning. Furthermore, large difference of $h_{max}$ position for single crystals with phosphorus doping compared to samples with doping instead of Ba or Fe clearly shows very different pinning mechanism in BaFe$_2$As$_{2-x}$P$_x$ single crystals versus BaFe$_{2-x}$M$_x$As$_2$ (M=Fe, Ni) and Ba$_{1-x}$A$_x$Fe$_2$As$_2$ (A=K, Na) samples where $h_{max}$ located around 0.3-0.4 allowing to suggest dominating NPP mechanism.

As one can see from Fig.5, in striking contrast to H//c-axis orientation, the $f_p(h)$ curves obtained at different temperatures in H//ab-plane configuration reach a maximum at different $h_{max}$ positions, thus demonstrating the absence of scaling. This result does not look very surprising since shielding current in H//ab-plane orientation consists of two components: current flowing along the ab-planes and current directed parallel to the c-axis. These two current components may be controlled by two different pinning mechanisms with different field and temperature dependence, thus, resulting in the absence of scaling.

Comparing values of the critical current density found at the same field in two field orientations, $J_c^{(c)}$ with $H//c$-axis and $J_c^{(ab)}$ with $H//ab$-plane, one can see that for samples BaFe$_{1.9}$Ni$_{0.1}$As$_2$ and BaFe$_{1.86}$Ni$_{0.14}$As$_2$, $J_c^{(c)}$ much exceeds $J_c^{(ab)}$ in almost all region of the vortex solid on the magnetic phase diagram (except small area close to irreversibility line $H_{irr}(T)$). Close to $H_{irr}(T)$ when pinning strength decreases, $J_c^{(ab)}(H)/J_c^{(c)}(H)$ ratio starts to grow rapidly, approaching $H_{irr}^{(ab)}/H_{irr}^{(c)}$ $vs$ $T$ ratio: this can be seen from $J_c^{(ab)}(3.6T)/J_c^{(c)}(3.6T)$ and $J_c^{(ab)}(2.3T)/J_c^{(c)}(2.3T)$ ratios for BaFe$_{1.9}$Ni$_{0.1}$As$_2$ and BaFe$_{1.86}$Ni$_{0.14}$As$_2$ samples respectively. Figure 6 where we show ratio $H_{irr}^{(ab)}/H_{irr}^{(c)}$ $versus$ $T$ for both our crystals (which as mentioned earlier roughly reflects superconducting anisotropy $\gamma = H_{c2}^{(ab)}/H_{c2}^{(c)} > 1$ in our samples) and critical current density ratio $J_c^{(ab)}/J_c^{(c)}$ $versus$ $T$ taken at several fields well illustrates this result. Our observation of an apparent inversion of the critical current density anisotropy finds a support within recently developed phenomenological approach to the effect of the specific origin of the pinning potential dependent on type of defects and impurities presenting in superconducting materials on the critical current anisotropy[24].

However, we cannot exclude another much more trivial possibility. In Ref.24, apparent inversion of critical current anisotropy was calculated for the in-plane current, while in our measurements in $H//ab$-plane geometry we probe shielding current that consists of two components with one of them flowing along the $ab$-plane and another directed parallel to the $c$-axis. If critical current component directed along the $c$-axis is much smaller compared to the in-plane critical current component one can observe strongly reduced shielding current in $H//ab$-plane orientation compared to $H//c$-axis geometry imitating apparent inversion of the critical current density anisotropy considered in van der Beek *et al* model.[24] Clearly, further studies of field behavior of independent in-plane and out-of-plane critical current components are needed to determine the critical current anisotropy in FBS single crystals of the 122 family.

**Summary**

In summary, we have measured isothermal irreversible magnetization loops of nearly optimally doped ($x$=0.1, $T_c \approx$ 19.5K) and slightly overdoped ($x$=0.14, $T_c \approx$ 13K) BaFe$_{2-x}$Ni$_x$As$_2$ single crystals in magnetic fields up to 14T applied perpendicular and parallel to the *ab*-plane. For both samples in two field orientations we observed critical current density exceeding $10^6$ A/cm$^2$ at low temperature suggesting strong intrinsic pinning in these samples. For both samples and for H//*c*-axis field orientation, the curves of normalized pinning force $f_p = F_p/F_p^{max}$ vs $h = H/H_{irr}$, measured at different temperatures fall in a single curve with peak position $h_{max} \approx$ 0.33 for BaFe$_{1.86}$Ni$_{0.14}$As$_2$ crystal and $h_{max} \approx$ 0.4 for BaFe$_{1.9}$Ni$_{0.1}$As$_2$ sample indicating single dominating normal point pinning mechanism. In the H//*ab*-planes geometry where shielding current consists of two components parallel and perpendicular to the c-axis $f_p(h)$ curves show no scaling.

**Acknowledgement**

This work was supported by the Russian Foundation for Basic Research (grant N 10-02-00680), by the Ministry of Education and Science of the Russian Federation and was performed using facilities of the International Laboratory of High Magnetic Fields and Low Temperatures and the Shared Research Facilities Center at P. N. Lebedev Physical Institute.

**TABLE 1**

Data for single crystals of 122 family of different compositions with around optimal doping level obtained by various groups: the superconducting critical temperature $T_c$, the slope of the upper critical field in dependence on temperature $dH_{c2}/dT$, zero field critical current density $J_c$ calculated from $M(H)$ curves measured with $H//c$-axis at temperatures shown in the Table, and normalized pinning force peak position $h_{max}=H_{max}/H_{irr}$.

| composition | $T_c$, K | $dH_{c2}/dT$, T/K | $J_c$, A/cm$^2$ | $h_{max}=H_{max}/H_{irr}$ | reference |
|---|---|---|---|---|---|
| BaFe$_{1.9}$Ni$_{0.1}$As$_2$ | 19.5 | -4.2 | 2.8x10$^6$ (at 4.2K) | 0.4 | this work |
| BaFe$_2$As$_{1.36}$P$_{0.64}$ | 28 | - | 4x10$^5$ (at 15.4K) | 0.7 | [5] |
| BaFe$_{1.84}$Co$_{0.16}$As$_2$ | 24.1 | -8 | 9x10$^5$ (at 4 K) | - | [6] |
| Na$_{0.75}$Ca$_{0.25}$Fe$_2$As$_2$ | 33.4 | | 1.1x10$^6$ (at 5 K) | - | [7] |
| BaFe$_{1.86}$Co$_{0.14}$As$_2$ | 22 | - | 2.6x10$^5$ (at 5 K) | - | [8] |
| BaFe$_{1.8}$Co$_{0.2}$As$_2$ | 24 | -1.7 | 6x10$^5$ (at 5 K) | - | [9] |
| BaFe$_{1.8}$Co$_{0.2}$As$_2$ | 22 | -2.5 | 4x10$^5$ (at 4.2 K) | 0.45 | [10] |
| BaFe$_{1.9}$Ni$_{0.1}$As$_2$ | 17.6 | -4.2 | 4x10$^5$ (at 2 K) | - | [11] |
| Ba$_{0.72}$K$_{0.28}$Fe$_2$As$_2$ | 32 | -4.4 | 3x10$^5$ (at 7 K) | - | [12] |
| Ba$_{0.65}$Na$_{0.35}$Fe$_2$As$_2$ | 29.4 | - | 1x10$^6$ (at 5 K) | 0.28 | [13] |
| Ba$_{0.68}$K$_{0.32}$Fe$_2$As$_2$ | 38.5 | -3.4 | 1.1x10$^6$ (at 10 K) | 0.43 | [14] |
| BaFe$_{1.85}$Co$_{0.15}$As$_2$ | 24.5 | -2.0 | 4.2x10$^5$ (at 10 K) | 0.37 | [14] |
| BaFe$_{1.91}$Ni$_{0.09}$As$_2$ | 18.5 | -2.2 | 2.3x10$^5$ (at 10 K) | 0.32 | [14] |

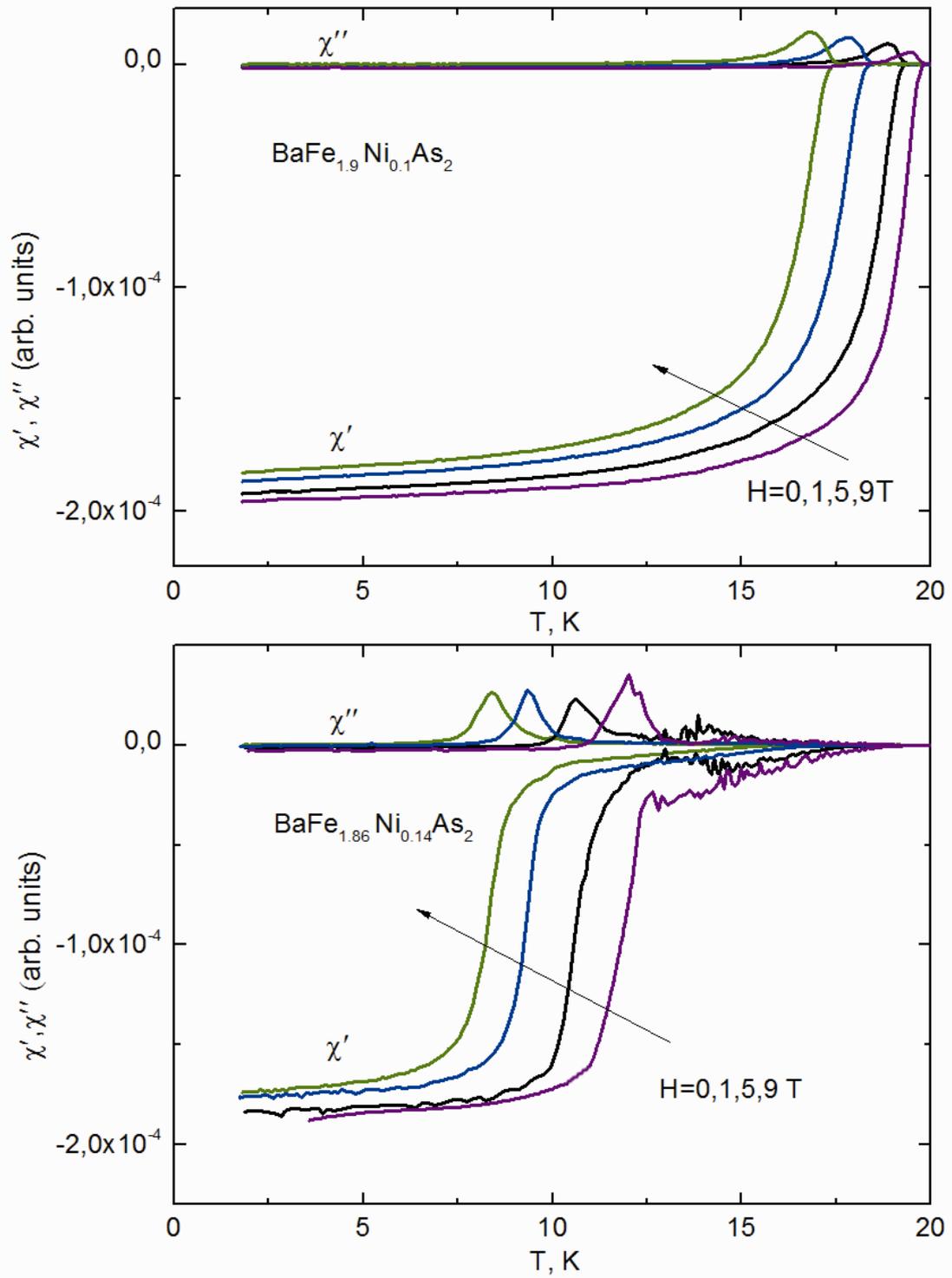

**Fig.1.** Pervakov et al

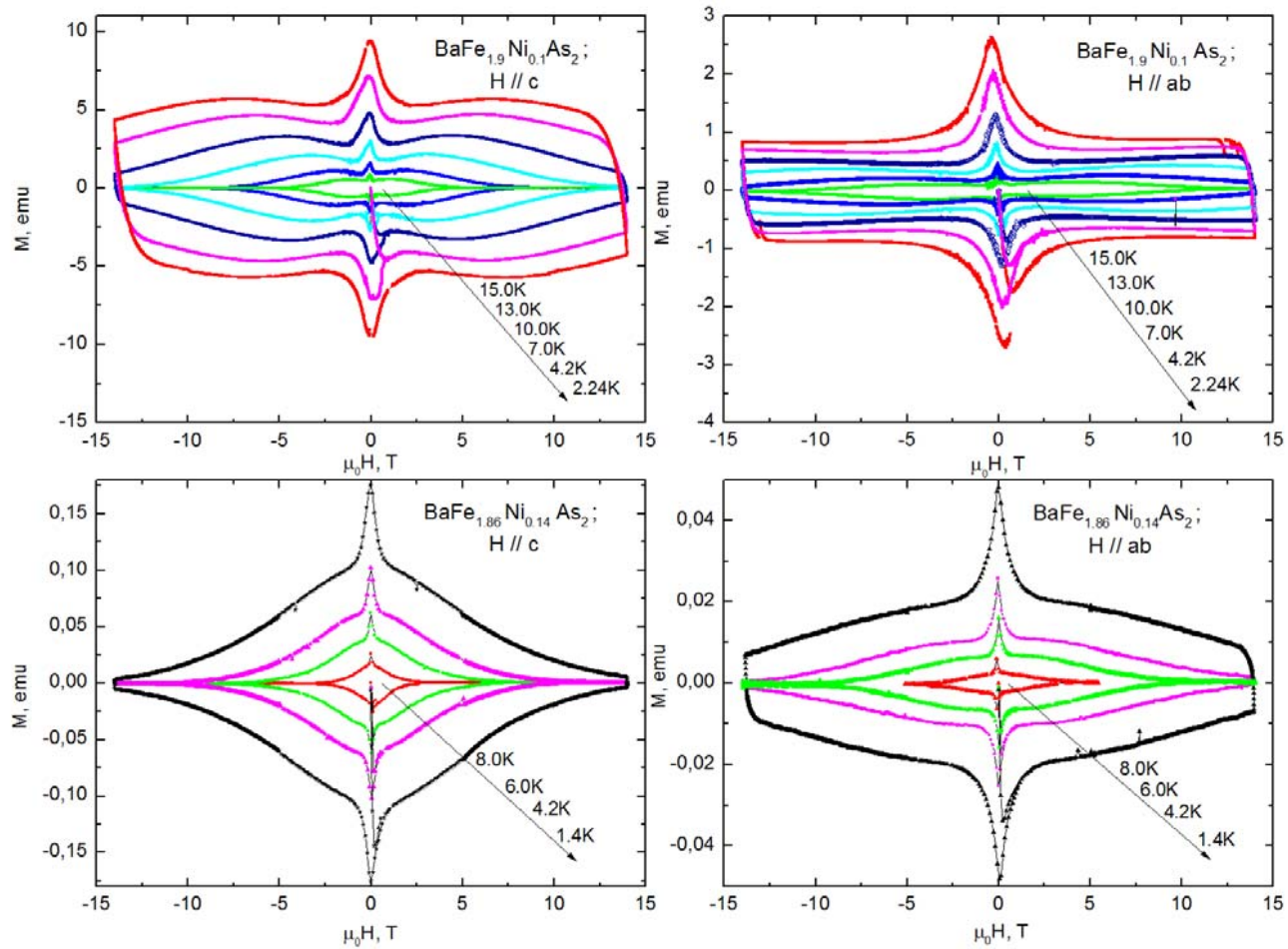

**Fig.2.** Pervakov et al

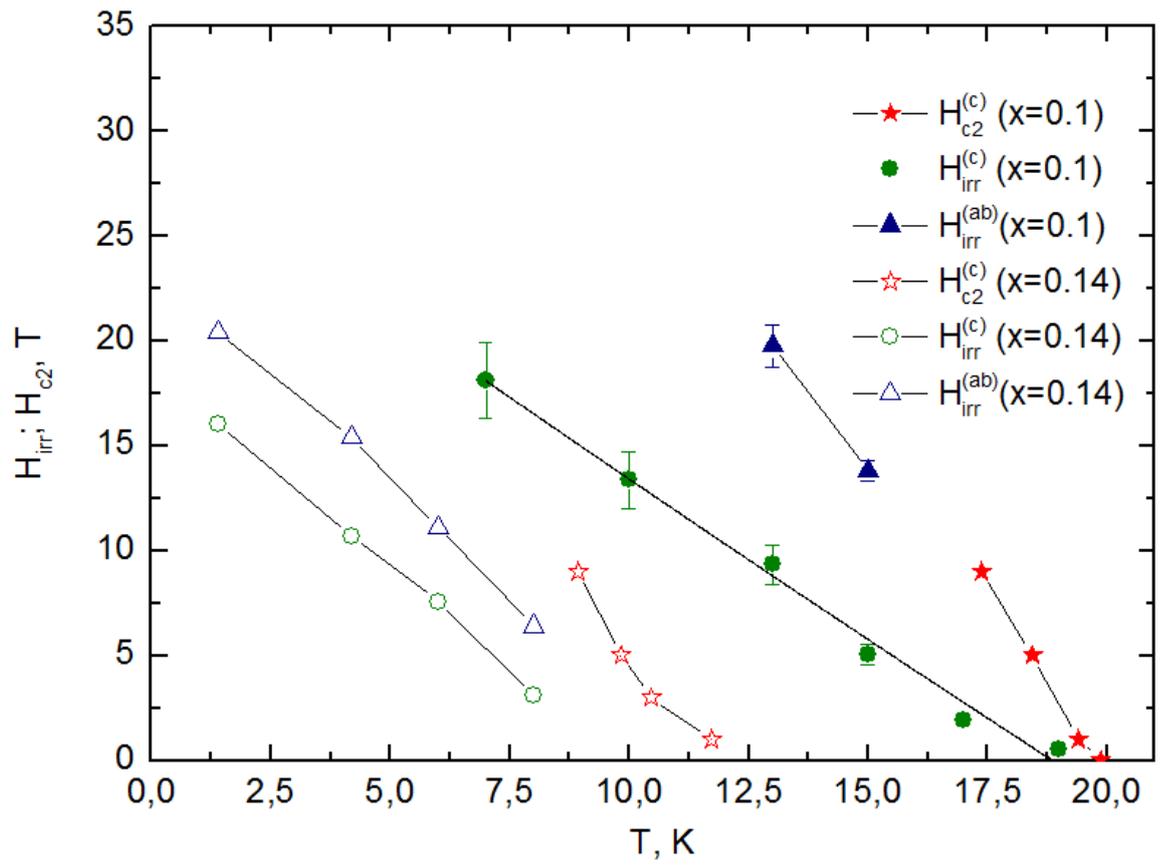

**Fig.3.** Pervakov et al

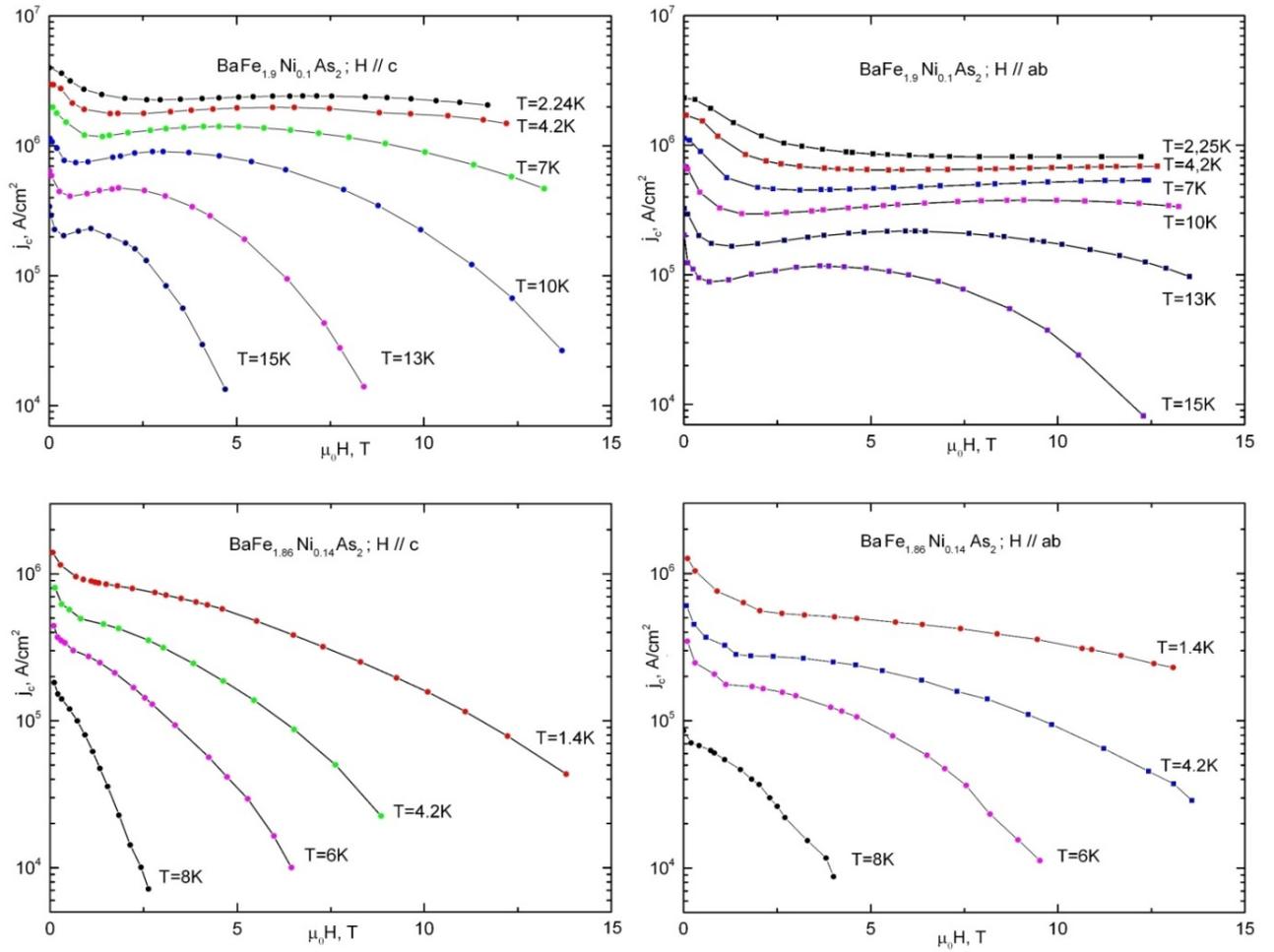

**Fig.4.** Pervakov et al

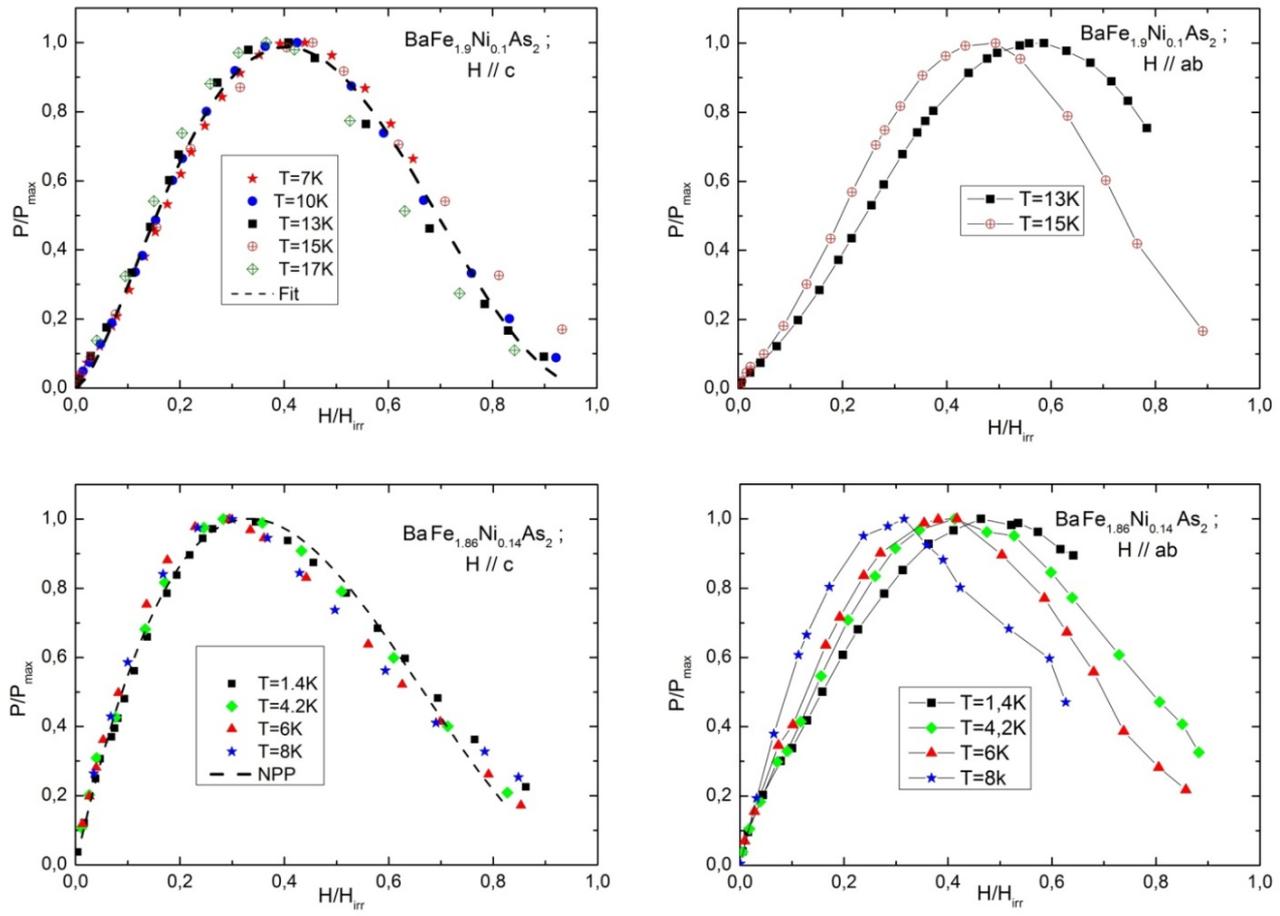

**Fig.5.** Pervakov et al

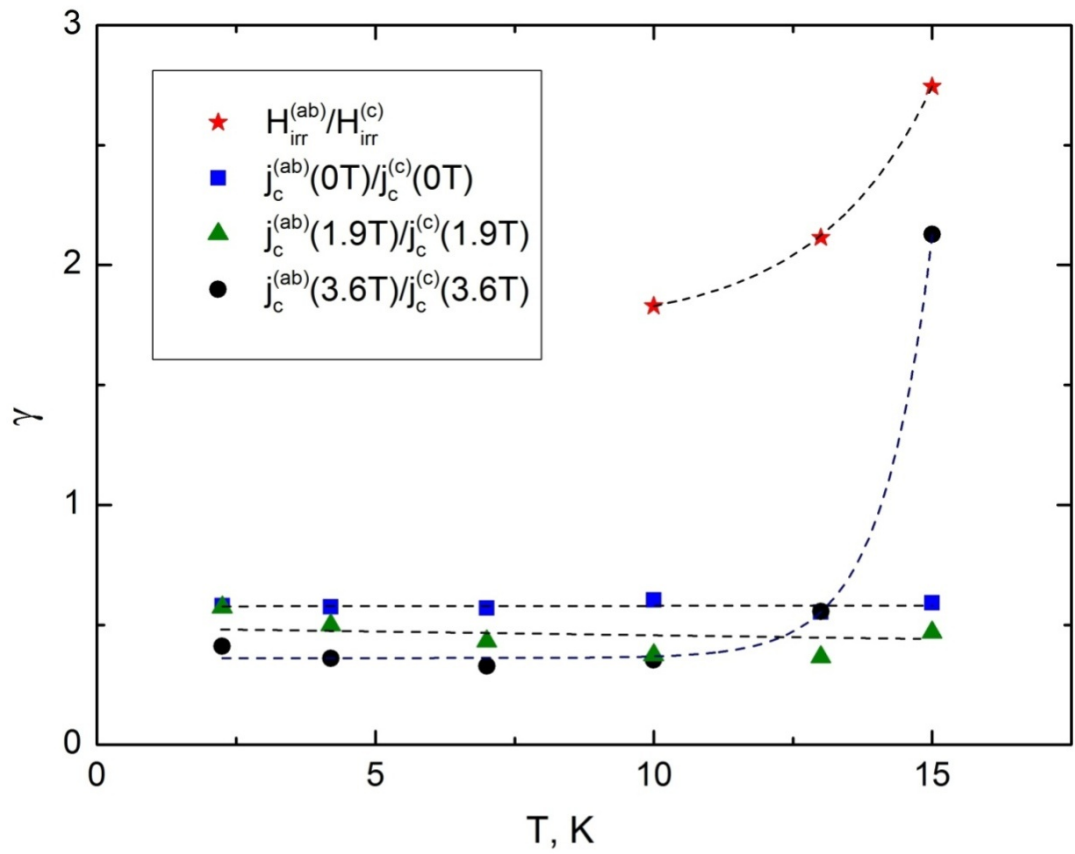
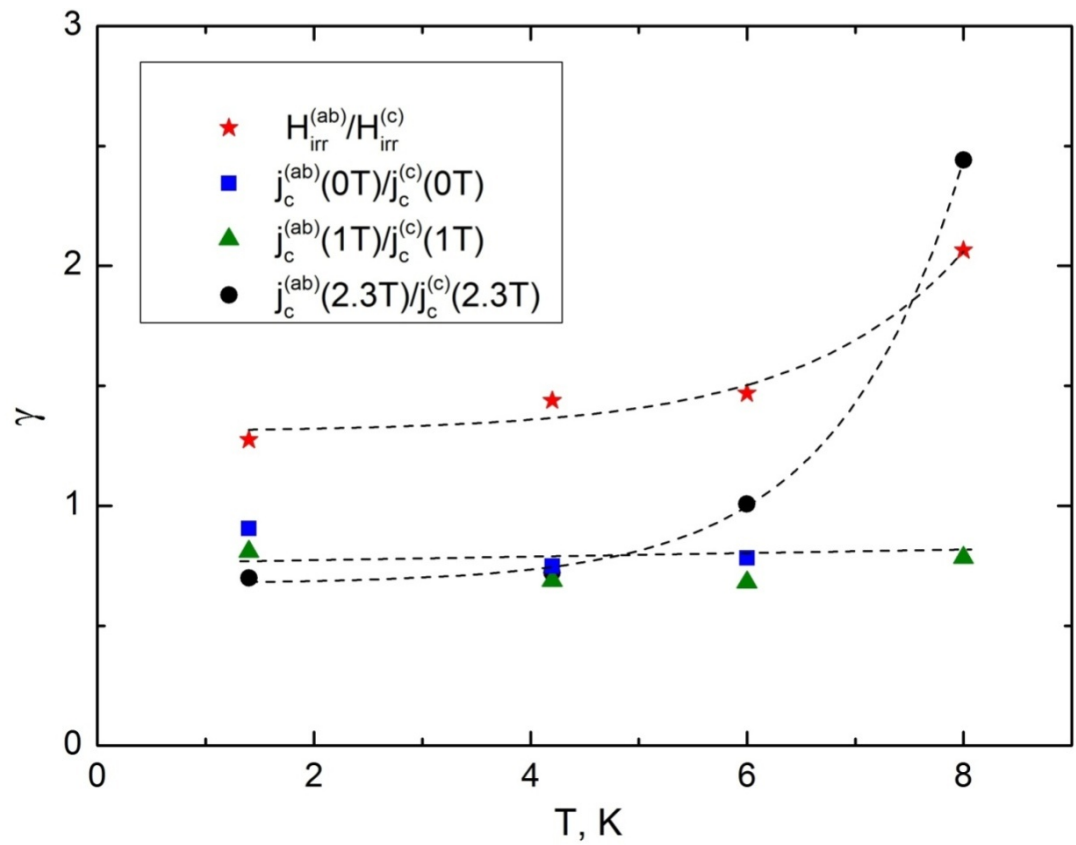

**Fig.6.** Pervakov et al

**Fugure captions**

**Fig.1.** Temperature dependence of the real, $\chi'$, and imaginary, $\chi''$, part of ac-susceptibility for BaFe$_{1.9}$Ni$_{0.1}$As$_2$ (top panel) and BaFe$_{1.86}$Ni$_{0.14}$As$_2$ single crystals (bottom panel) in magnetic fields of 0, 1, 5 and 9 T (from right to the left) applied along the *c*-axis.

**Fig.2.** Irreversible isothermal magnetization loops for BaFe$_{1.9}$Ni$_{0.1}$As$_2$ (top panels) and BaFe$_{1.86}$Ni$_{0.14}$As$_2$ (bottom panels) crystals measured at various temperatures as indicated in the figure. Magnetic field applied along the *c*-axis (left panels) and along the *ab*-plane (right panels).

**Fig.3.** Magnetic phase diagram for BaFe$_{1.9}$Ni$_{0.1}$As$_2$ (closed symbols) and BaFe$_{1.86}$Ni$_{0.14}$As$_2$ (opened symbols) samples. Lines are a guide to the eye.

**Fig.4.** Field dependence of critical current density for BaFe$_{1.9}$Ni$_{0.1}$As$_2$ and BaFe$_{1.86}$Ni$_{0.14}$As$_2$ samples extracted from *M(H)* bulk magnetization curves measured in two magnetic field orientations at different temperatures as shown in the figure. Lines are a guide to the eye.

**Fig.5.** Normalized pinning force $f_p = F_p/F_p^{max}$ vs reduced field $H/H_{irr}$ obtained at different temperatures as shown in the figure for BaFe$_{1.9}$Ni$_{0.1}$As$_2$ and BaFe$_{1.86}$Ni$_{0.14}$As$_2$ single crystals in two field configurations. Dotted line in the lower left panel shows approximation of experimental data by $f(h) \sim h(1-h)^2$ NPP expression. Dotted line in the upper left panel displays the best fit to $f(h) \sim h^{1.59}(1-h)^{2.26}$ expression. Lines in top and bottom right panels are a guide to the eye.

**Fig.6.** Temperature dependence of apparent anisotropy of critical current density $J_c^{(ab)}(H)/J_c^{(c)}(H)$ taken at a few different fields as shown in the figure and irreversibility field ratio $H_{irr}^{(ab)}/H_{irr}^{(c)}$ for BaFe$_{1.9}$Ni$_{0.1}$As$_2$ (top panel) and BaFe$_{1.86}$Ni$_{0.14}$As$_2$ single crystals (bottom panel). Dotted lines in both panels are a guide to the eye.